\documentstyle[12pt,axodraw,epsfig,prc,aps]{revtex}
\newcommand{\fslash}[1]{\mbox{$\!\not\!#1$}}

\begin{document}

\draft
\title{Massive Meson Fluctuation in NJL Model}

\author{Mei Huang$^{1}$, Pengfei Zhuang$^{2}$, Weiqin Chao$^{1,3}$ \\ 
{ \small $^1$ Institute of High Energy Physics, Chinese Sciences Academy, Beijing 100039, China\\
  $^2$ Physics Department, Tsinghua University, Beijing 100084, China\\
  $^3$ CCAST, Beijing 100080, China}}
\date{\today}

\maketitle

\vskip 1cm

\begin{abstract}
Based on the self-consistent scheme beyond mean-field approximation in the 
large $N_c$ expansion, including current quark mass explicitly,
a general
scheme of SU(2) NJL model is developed. To ensure the quark self-energy
expanded in the proper order of $N_c$, an approximate internal
meson propagator is deduced, which is in order of $O(1/N_c)$.
In our scheme, adopting the method of external momentum expansion,
all the Feynman diagrams are calculated 
in a unified way by only expanding the quark propagator. 
Our numerical results show, that different from the mean field approximation
in which the explicitly chiral symmetry breaking is invisible, 
the effect of finite pion mass can be seen clearly 
when beyond mean-field approximation.
\end{abstract}

\newpage

\section{Introduction}
As we all know, chiral symmetry breaking was originally explained quite well by 
Nambu-Jona-Lasinio 
(NJL) model as early as 1961 \cite{Nambu}. Like in superconductivity, the strong
attractive force between quark and antiquark in the $J^P=0^+$ channel can create
non-perturbative ground state with ${\bar q}q$ condensation. Due to the pair condensation, 
the original symmetry between massless left and right-handed quarks is broken 
down to $U_V(N_f)$, and then the quarks obtain constituent mass. 
The remaining residual interactions between the constituent quarks bind them into collective 
excitations, i.e., hadrons in the chiral symmetry breaking vacuum.
Especially in the pseudoscalar channel the residual strong 
interaction creates massless pions as Goldstone bosons in the chiral limit.
When a small current quark mass $m_0$ is introduced in the theory, chiral 
symmetry is explicitly broken,
and pion obtains its small physical mass $m_{\pi}$.

Although the NJL model has two serious drawbacks, i.e., lacks of
confinement and renormalizability,
it is still regarded as an applicable model at low momentum, 
especially for dealing with
processes of pion, such as pion-pion scattering near threshold.  

Traditionally, the scheme of the 
NJL model is represented by two Schwinger-Dyson (SD) equations, one 
is for 
the constituent quark propagator, and the other is for the composite 
meson propagator. At the lowest level, the applications of the NJL model  
are based upon mean-field approximation \cite{NJL1} - \cite{m03}, 
 i.e., Hartree approximation to 
the gap equation
for quark mass and the random-phase approximation (RPA) to the Bethe-Salpeter 
equation for meson mass.  
It is clear, that at this level the solution of the gap equation 
determines the meson propagators, but the solution of meson SD equation has 
no feedback to
the quark propagator. 
Since the constituent quark propagator is the fundamental element, from which all 
the quantities,
including quark mass, meson masses and quark-antiquark condensate, are calculated, 
it is necessary to consider the back contribution of meson modes to the 
quark propagator.  

Among efforts \cite{cao} - \cite{akama} to go beyond the mean-field approximation, Refs.
\cite{NPA} and \cite{Ann} are in a  
chirally symmetric self-consistent approximation, namely
the chiral properties such as the Goldstone's theorem, the Goldberger-Treiman 
relation and
the conservation of the quark axial current are exactly preserved in the chiral 
limit of the NJL model. 
By using effective action method in a semi-bosonized way, and 
expanding the action to one quark-loop and one-meson-loop in \cite{NPA},
or directly evaluating the 
Feynman diagrams under the constraint to keep the chiral relations at quark level
in \cite{Ann}. 

In this paper, we extend the method of \cite{Ann} to a general scheme
with explicit chiral symmetry breaking in the SU(2) NJL model. 
Different from the case in the chiral limit, we must be careful 
to deal with the form
of internal meson propagators.
In a way different from \cite{Ann},
we regard the constituent quark as the fundamental element and only expand quark's
propagator in the power of small external momentum in the calculation of  
Feynman diagrams. 

In the process to go beyond the mean-field approximation, we have to deal 
with the divergent integrals
of quark loops and meson loops. We adopt Pauli-Villars regulation 
\cite{NJL2,berna} to treat
divergent integrals resulted from quark loops, 
and choose a covariant cutoff $\Lambda_b$ for the meson momentum.     
There are four parameters in our treatment, namely the current quark mass $m_0$, 
quark coupling constant $G$, 
fermionic cut-off $\Lambda_f$ and bosonic cut-off $\Lambda_b$, to be fixed. 
In the mean-field approximation, the three parameters $m_0$, $G$, 
$\Lambda_f$ are usually fixed by comparing with the pion mass $m_{\pi}=140$ MeV, 
pion decay constant $f_{\pi}=92.4$ MeV
and the quark condensate $1/2<{\bar q}q>^{1/3}=-250$ MeV. In the near 
future, the DIRAC experiment will
measure the $\pi-\pi$ scattering lengths in good precision, which will 
shed some light
on the quark condensate \cite{dirac}.
To see clearly the quark condensate dependence of the four parameters,
 we give only the quark condensate a reasonable 
constraint: -300 MeV $\sim $ -200 MeV. 

The outline of this paper is as follows: In section 2, we briefly review 
the general scheme represented by 
two Schwinger-Dyson equations in the SU(2) NJL model. In Section 3, we introduce 
the method of external momentum expansion, and prove
a general relation between the pion polarization function and the axial-vector 
matrix element. 
We also deduce the 
internal meson propagator to $O(1/N_c)$ order in the $N_c$ expansion. 
Our numerical results with 
mesonic contributions and the effect of explicit chiral symmetry breaking will be
shown in section 4. The conclusions are given at the end. 

\section{ The Fundamental Scheme in SU(2) NJL Model}
\subsection{Two Schwinger-Dyson equations in SU(2) NJL model}
~In this section, we briefly review the traditional scheme of SU(2) NJL model with large
$N_c$ expansion. 
The two-flavor NJL model is defined through the Lagrangian density,
\begin{eqnarray}
\label{lagr}
{\cal L} = \bar{\psi}(i\gamma^{\mu}\partial_{\mu}-m_0)\psi + 
  G[(\bar{\psi}\psi)^2 + (\bar{\psi}i\gamma_5{\bf {\vec \tau}}\psi)^2 ],
\end{eqnarray}
here $G$ is the effective coupling constant of dimension ${\rm GeV}^{-2}$, and
$m_0$ is the current quark mass, assuming isospin degeneracy of the 
$u$ and $d$ quarks,
and $\psi, \bar{\psi}$ are quark fields with flavor, colour and spinor indices
suppressed. 
  
  The traditional non-perturbative method of NJL model is inspired 
from many-body theory.  
The complete description is represented by two Schwinger-Dyson (SD) integral equations, i.e.,
the constituent quark propagator, see Fig. 1a, and the composite meson propagator, see
Fig. 1b.  
\begin{center}
\begin{picture}(350,100)(0,0)
\SetWidth{2}
\Text(60,80)[]{a}
\Line(50,60)(90,60)
\Line(200,60)(220,60)
\GOval(200,85)(10,8)(0){0.8}
\SetWidth{1}
\Text(102,60)[]{=}
\Line(110,60)(150,60)
\Text(165,60)[]{+}
\Line(180,60)(200,60)
\DashLine(200,60)(200,75){4}
\Vertex(200,60){2.5}
\Vertex(200,75){2.5}

\SetWidth{2}
\Text(60,40)[]{b}
\DashLine(55,20)(90,20){4}
\Vertex(55,20){2.5}
\Vertex(90,20){2.5}
\DashLine(250,20)(270,20){4.5}
\GOval(240,20)(10,8)(90){0.8}
\SetWidth{1}
\Text(110,20)[]{=}
\Vertex(130,20){2.5}
\Vertex(165,20){2.5}
\DashLine(130,20)(165,20){4}
\Text(180,20)[]{+}
\DashLine(205,20)(230,20){4}
\Vertex(205,20){2.5}
\Vertex(230,20){2.5}
\Vertex(250,20){2.5}
\Vertex(270,20){2.5}
\end{picture}
\end{center}
\begin{figure}
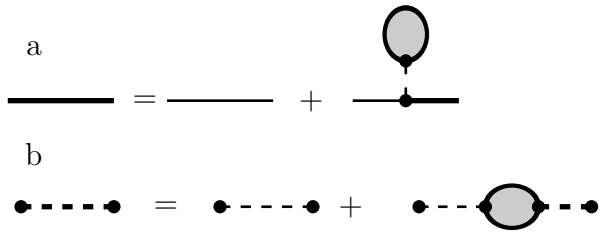

\begin{center}
\caption[]{ The two Schwinger-Dyson equations for the quark propagator $a$ and   
the meon propagator $b$. The light dashed lines 
in $a$ and $b$ represent the four-fermion vertex
$2iG$, and the one-vertex gray bubble in $a$ and the two-vertex gray bubble in $b$ 
indicate the quark self-energy and the meson polarization function respectively.}
\end{center}
\end{figure}
\vspace*{-1truecm}
The two SD equations must couple to 
each other self-consistently and keep chiral symmetry in the chiral limit.  
The one-vertex grey bubble $kernel ~a$ in Fig. 1$a$ represents the quark self-energy,
and the two-vertex grey bubble $kernel ~b$ in Fig. 1$b$ indicates the meson polarization function.
In the case of SU(2), meson
modes refer to only pseudoscalar mesons $\pi$ and scalar mesons $\sigma$.
 
It is difficult to give the full expressions of $kernel ~a$ and $kernel ~b$ in Fig. 1.
Usually an approximation scheme called large $N_c$ expansion is adopted in
NJL model, i.e., the both kernels expanded in $1/N_c$, where $N_c$ is the number of color.
In this scheme, $GN_c$ is a constant when $N_c \rightarrow \infty$, so coupling constant
scales like $G \sim N_c^{-1}$. One can find the detailed description of large $N_c$ expansion in
\cite{witten,equack}. 
V.Dmitra{\v s}inovi{\' c} et.al. proved in their paper \cite{Ann} that the $kernel ~a$ and $kernel ~b$ 
shown in Fig. 2 are self-consistent leading and subleading order
in $N_c$ expansion and can keep all the chiral relations
in the chiral limit. 
 
The leading $O(1)$ order of quark self-energy $kernel ~a$ in Fig. 2 
is named $m_H$ after Hartree approximation, and the subleading $O(1/N_c)$ order
is called $\delta m$. 
Including current quark mass $m_0$, 
the gap equation for quarks can be expressed as
\begin{eqnarray}
\label{gap}
m = m_0 + m_H +\delta m, 
\end{eqnarray}
where $m_H=16imGN_cF$, and $F$ is given in Appendix in order to compactly discuss physics
in the text.

The leading $O(N_c)$ and subleading $O(1)$ order of meson polarization function $kernel ~b$
are expressed as $\Pi^{(RPA)}_{M}(k)$ and $\delta \Pi_{M}(k)$ respectively,
here $M$ represents $\pi$ or $\sigma$.
It is clear to see that the back interaction which conserves all the chiral 
properties is reflected in the contribution from the meson propagator to 
the quark mass. 
The total meson propagator can be expressed in terms of the total polarization function
\begin{eqnarray}
\label{prog}
-{\rm i} D_{M}(k) & = & \frac{2iG}{1-2G \Pi_{M}(k)}, 
\end{eqnarray}
where $\Pi_{M}(k) = \Pi^{(RPA)}_{M}(k) + \delta \Pi_{M}(k)$, and the RPA part is
$\Pi^{(RPA)}_{\pi}(k)=8iN_cF-4iN_ck^2I(k)$ for $\pi$ and 
$\Pi^{(RPA)}_{\sigma}(k)=8iN_cF-4iN_c(k^2-4m^2)I(k)$ for $\sigma$, 
and the function $I(k)$ is defined in Appendix. 
\begin{center}
\begin{picture}(450,100)(0,0)
\SetWidth{2}
\GOval(30,50)(15,10)(0){0.8}
\Vertex(30,35){2.5}
\CArc(120,50)(15,0,360)
\Vertex(120,35){2.5}
\Text(30,19)[]{$kernel~a$}
\Text(120,19)[]{$m_H$}
\CArc(200,50)(15,0,360)
\DashCArc(200,70)(15,320,220){4}
\SetWidth{1}
\Text(75,50)[]{=}
\Text(160,50)[]{+}
\Vertex(200,35){2.5}
\Vertex(189,61){2.5}
\Vertex(210,61){2.5}
\Text(200,19)[]{$\delta m$}
\Text(200,75)[]{$\pi$,$\sigma$}
\end{picture}
\begin{picture}(450,100)(0,0)
\SetWidth{2}
\GOval(30,50)(15,10)(90){0.8}
\Text (30,15)[]{$kernel ~b$}
\Vertex(15,50){2.5}
\Vertex(45,50){2.5}
\Text(67.5,50)[]{=}
\CArc(105,50)(15,0,360)
\Vertex(90,50){2.5}
\Vertex(120,50){2.5}
\Text(110,15)[]{$\Pi^{(RPA)}_{M}(k)$}
\Text(135.5,50)[]{+}
\CArc(180,50)(15,0,360)
\DashLine(180,65)(180,35){4}
\Vertex(165,50){2.5}
\Vertex(180,65){2.5}
\Vertex(180,35){2.5}
\Vertex(195,50){2.5}
\Text(180,15)[]{$\delta \Pi^{(b)}_{M}(k)$}
\Text(180,50)[]{$\pi$,$\sigma$}
\Text(215.5,50)[]{+}
\CArc(260,50)(15,0,360)
\DashCArc(260,70)(15,320,220){4}
\Vertex(249,60){2.5}
\Vertex(248,41){2.5}
\Vertex(271,60){2.5}
\Vertex(272,41){2.5}
\Text(260,75)[]{$\pi$,$\sigma$}
\Text(260,15)[]{$\delta \Pi^{(c)}_{M}(k)$}
\Text(295.5,50)[]{+}
\SetWidth{2}
\CArc(340,50)(10,0,360)
\CArc(400,50)(10,0,360)
\Vertex(330,50){2.5}
\Vertex(340,60){2.5}
\Vertex(340,40){2.5}
\Vertex(410,50){2.5}
\Vertex(400,60){2.5}
\Vertex(400,40){2.5}
\Text(370,65)[]{$\pi$}
\Text(370,35)[]{$\sigma$}
\Text(370,15)[]{$\delta \Pi^{(d)}_{M}(k)$}
\DashCurve{(340,60)(360,70)(380,70)(400,60)}{3.2}
\DashCurve{(340,40)(360,30)(380,30)(400,40)}{3.2}
\end{picture}
\end{center}
\begin{figure}
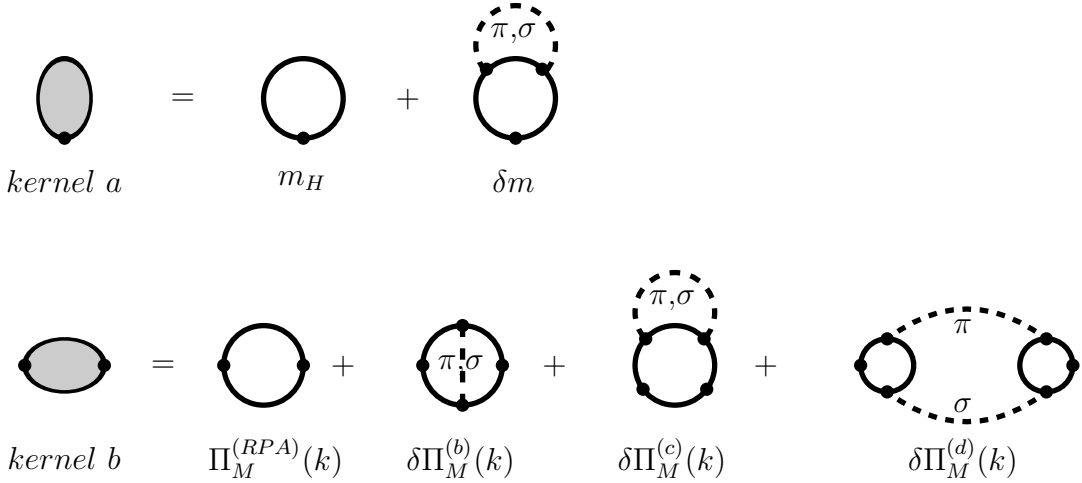

\begin{center}
\caption[]{ $kernel ~a$ and $kernel ~b$ in the quark and meson propagators. 
$m_H$ and $\delta m$ are the leading and subleading contributions to 
the quark mass. $\Pi^{(RPA)}_M$ and $\delta \Pi^{(b,c,d)}_M$ are the leading 
and subleading order contributions to meson polarization function. The heavy solid lines
indicate the constituent quark propagator,  
and the heavy dashed lines represent 
$\pi$ or $\sigma$ propagator $-{\rm i} D^{(RPA)}_{M}(q)$ in RPA approximation. }
\end{center}
\end{figure}
\vspace*{-1truecm} 
In the NJL model, mesons are bound states of constituent quark and
anti-quark, the meson mass $m_{M}$ satisfies the meson propagator's pole condition
\begin{eqnarray}
\label{pole}
1-2G\Pi_{M}(k^2=m_{M}^2)=0,
\end{eqnarray}
and the coupling constant $g_{M qq}$ is determined by the residue at the pole
\begin{eqnarray}
\label{couple}
g_{M qq}^{-2} = (\partial \Pi_{M}(k)/\partial k^2)^{-1}|_{k^2=m_{M}^2}.
\end{eqnarray}
Performing direct calculation of the Feynman diagrams in Fig. 2 leads to the
explicit expressions for the quark mass
\begin{eqnarray}
- {\rm i}\delta m & = & 12GN_c\int\frac{d^4q}{(2\pi)^4}[-{\rm i}D^{(RPA)}_{\pi}(q)]
              \int\frac{d^4p}{(2\pi)^4} {\rm Tr}[\gamma_5 S(p+q) \gamma_5 S(p) 1 S(p)]  \nonumber \\
         & - & 4 GN_c \int\frac{d^4q}{(2\pi)^4}[-{\rm i}D^{(RPA)}_{\sigma}(q)]
         \int\frac{d^4p}{(2\pi)^4}{\rm Tr}[1 S(p+q) 1 S(p) 1 S(p)],
\end{eqnarray}
and for pion polarization fluctuations
\begin{eqnarray}
&\delta \Pi^{(b)}_{\pi}(k)& = 2iN_c[\int\frac{d^4q}{(2\pi)^4}[-{\rm i}D^{(RPA)}_{\pi}(q)]
                \int\frac{d^4p}{(2\pi)^4}
           {\rm Tr}[\gamma_5 S(p+q) \gamma_5 S(p)\gamma_5 S(p-k) \gamma_5 S(p-k+q)] \nonumber \\ 
& & + \int\frac{d^4q}{(2\pi)^4}[-{\rm i}D^{(RPA)}_{\sigma}(q)]
                \int\frac{d^4p}{(2\pi)^4}  
       {\rm Tr}[\gamma_5 S(p+q) 1 S(p) \gamma_5 S(p-k) 1 S(p-k+q)]], \\
&\delta \Pi^{(c)}_{\pi}(k) & = -12iN_c[\int\frac{d^4q}{(2\pi)^4}[-{\rm i}D^{(RPA)}_{\pi}(q)]
                \int\frac{d^4p}{(2\pi)^4}  
         {\rm Tr}[\gamma_5 S(p) \gamma_5 S(p+q) \gamma_5 S(p) \gamma_5 S(p-k)] \nonumber \\
& & -  {1 \over 3} \int\frac{d^4q}{(2\pi)^4}[-{\rm i}D^{(RPA)}_{\sigma}(q)]
                \int\frac{d^4p}{(2\pi)^4}  
      {\rm Tr}[\gamma_5 S(p) 1 S(p+q) 1 S(p) \gamma_5 S(p-k)]], \\
&\delta \Pi^{(d)}_{\pi}(k) & = i \int\frac{d^4q}{(2\pi)^4}[-{\rm i} \Gamma^{\sigma \pi \pi}(k,q)]
                  [-{\rm i}D^{(RPA)}_{\pi}(q)] 
            [-{\rm i}D^{(RPA)}_{\sigma}(q-k)][-{\rm i} \Gamma^{\sigma \pi \pi}(q,k)].
\end{eqnarray}
Here, the expression $\Gamma^{\sigma \pi \pi}(k,q)$ is the ($\sigma \pi \pi$)-vertex 
\begin{eqnarray}
\label{vertex1}
-{\rm i} \Gamma^{\sigma \pi \pi}(k,q) = -4N_ci\int\frac{d^4p}{(2\pi)^4} 
               {\rm Tr}[\gamma_5 S(p+k) 1 S(p+q) \gamma_5 S(p)],
\end{eqnarray}
$S(p)=1/({\fslash p}-m)$ the quark propagator and  $-{\rm i} D^{(RPA)}_{\pi, \sigma}(q)$ the internal
meson propagator. In all the formulae of this paper, $k$ means the $\pi$ external momentum, $p$ the 
internal quark momentum and $q$ the internal meson momentum;

\subsection{Pion Decay Constant}
~Another important quantity is the pion decay constant $f_{\pi}$ which is calculated 
from the vacuum to one-pion axial-vector matrix element, replacing one vertex 
$i \gamma_5{\bf{\vec \tau}}$ of $kernel ~b$ in Fig. 2 by $ig_{\pi qq}(k)\gamma_5\gamma_{\mu}
{\bf{\vec \tau}}/2$. Similar to the expression of $\Pi_{\pi}(k)$, we express $f_{\pi}(k)$
as 
\begin{eqnarray}
f_{\pi}(k) & = & f_{\pi}^{(RPA)}(k) + \delta f_{\pi}(k) \nonumber \\
      & = & f_{\pi}^{(RPA)}(k) + \delta f_{\pi}^{(b)}(k) + \delta f_{\pi}^{(c)}(k) 
      + \delta f_{\pi}^{(d)}(k).
\end{eqnarray} 
Since the Feynman diagrams for $\Pi_{\pi}(k)$ and $f_{\pi}$ 
have the same structure, we expect to 
find a general relation between them. Starting from the axial-vector matrix element
$<0|J_{\mu 5}^{i}|\pi^{j}(k)>=if_{\pi}(k)k_{\mu}\delta^{ij}$,
and using the single-quark axial Ward identity involving the quark propagator $S(p)$,
\begin{eqnarray}
\gamma_5{\fslash k}=2m\gamma_5+\gamma_5S^{-1}(p+k)+S^{-1}(p)\gamma_5,
\end{eqnarray}
one can deduce the general relations 
functions  
\begin{eqnarray}
\label{fgpi}
\frac{k^2 f^{(RPA)}_{\pi}(k)}{g_{\pi qq}(k)} & = & m \Pi^{(RPA)}_{\pi}(k)
                                         -\frac{m_H}{2G}, \nonumber \\
\frac{k^2 \delta f_{\pi}(k)}{g_{\pi qq}(k)} & = & m \delta \Pi_{\pi}(k)
                                         -\frac{\delta m}{2G}.
\end{eqnarray}
The detailed proof can be found in Ref.\cite{Ann}.
Using pion's pole condition Eq.~(\ref{pole}) and gap equation Eq.~(\ref{gap}), we get finally
the relation
\begin{eqnarray}
\label{onshell}
\frac{m_{\pi}^2f_{\pi}}{g_{\pi qq}} = \frac{m_0}{2G},
\end{eqnarray}
where we have defined the on-shell quantities 
$f_{\pi}=f_{\pi}(k^2=m_{\pi}^2), \ \  g_{\pi qq}=g_{\pi qq}(k^2=m_{\pi}^2)$.

While in the chiral limit, $f_{\pi}$ satisfies the Goldberger-Treiman relation $f_{\pi}g_{\pi qq}=m$.

\section{External Momentum Expansion}
\subsection{Method of External Momentum Expansion}
~We are familiar with one-loop Feynman diagrams, but 
it is difficult to analytically calculate the two-loop Feynman diagrams such as those in $kernel ~b$.
Since pion's external momentum $k^2=m_{\pi}^2$ is small, we will calculate the subleading 
corrections in $kernel ~b$ to $k^2$ term only, 
but keep the complete 
contributions from the leading order.

In the NJL model, the constituent quarks are the fundamental elements, and the mesons are 
their bound. 
In fact, all the quantities in the NJL model are calculated from the
quark propagator $S(p)=1/({\fslash p}-m)$. Therefor, if we want to build
up a self-consistent expansion in the caculation of the the 2-loop diagrams,
we should only expand the quark propagator in the external 
momentum $k$
\begin{eqnarray}
\label{expand}
S(p \mp k) = \frac{1}{\fslash p \mp \fslash k-m}=S(p) \pm S(p){\fslash k}S(p)+
S(p){\fslash k}S(p){\fslash k}S(p) + \cdots .
\end{eqnarray}
With this expansion, all the two-loop diagrams can be expanded naturally
in $k$ in a unified way.

The Lorentz covariance gives the expansion form of $\Pi_{\pi}(k) $ as
\begin{eqnarray}
\label{expi}
\Pi_{\pi}(k) & = & \Pi^{(RPA)}_{\pi}(k) + \delta \Pi_{\pi}(k), \nonumber \\
\delta \Pi_{\pi}(k) & = & \delta \Pi_{\pi}(0) + k^2 \delta \Pi_{\pi}^{(2)}(0) + k^4 \delta \Pi_{\pi}^{(4)}(0)
                  + \cdots.
\end{eqnarray}
In the following, we consider  $ \delta \Pi_{\pi}(k)$ to the $k^2$ term only. 
In this approximation, the pole condition for the pion propagator is simplified as
\begin{eqnarray}
\label{expole}
1-2G(\Pi^{(RPA)}_{\pi}(k^2=m_{\pi}^2) + \delta \Pi_{\pi}(0) + 
      m_{\pi}^2 \delta \Pi_{\pi}^{(2)}(0)) = 0.
\end{eqnarray}
With the expression of $\Pi^{(RPA)}_{\pi}(k)$ and the gap equation Eq. ~(\ref{gap}) we 
obtain the gap equation to determine the pion mass,
\begin{eqnarray}
\label{mpi}
m_{\pi}^2 = \frac{m_0}{Gm(-8N_ciI(m_{\pi})+2 \delta \Pi_{\pi}^{(2)}(0))}. 
\end{eqnarray} 
Similarly, the expansion of the residue of the pole to the order $k^2$
gives the pion-quark coupling constant 
\begin{eqnarray}
\label{gpiqq}
g_{\pi qq}^{-2}  = (\partial \Pi^{(RPA)}_{\pi}(k)/\partial k^2)^{-1}|_{k^2=m_{\pi}^2} + 
                    \delta \Pi_{\pi}^{(2)}(0),
\end{eqnarray}
where $(\partial \Pi^{(RPA)}_{\pi}(k)/\partial k^2)^{-1} = -2N_ci(I(0)+I(k)-k^2K(k))$,        
and the quark-loop integral $K(k)$ is defined in Appendix. 

\subsection{Calculation of $\delta \Pi_{\pi}^{(2)}(0)$}
~It is tedious to calculate $\delta \Pi_{\pi}^{(2)}(0)$ directly. 
Here we will use another equivalent way by calculating the first term
of the axial-vector matrix element in the 
external momentum expansion. 

We have shown the general relations between pion polarization function and 
axial-vector matrix element in Eq.~(\ref{fgpi}). 
If $k^2=0$, i.e., in the chiral limit, Eq.~(\ref{fgpi}) becomes
\begin{eqnarray}
\label{limit}
\frac{m_H}{2G} = m \Pi^{(RPA)}_{\pi}(0), \ \ \ \frac{\delta m}{2G} & = & m \delta \Pi_{\pi}(0).
\end{eqnarray}
Thus the Eq. ~(\ref{fgpi}) in general case can be rewritten as
\begin{eqnarray}
\label{fgpi1}
\frac{k^2 f^{(RPA)}_{\pi}(k)}{g_{\pi qq}(k)} = m ( \Pi^{(RPA)}_{\pi}(k)- \Pi^{(RPA)}_{\pi}(0)), \ \
\frac{k^2 \delta f_{\pi}(k)}{g_{\pi qq}(k)} =  m ( \delta \Pi_{\pi}(k)-\delta \Pi_{\pi}(0)),
\end{eqnarray}
With the expansion form of $\Pi_{\pi}(k)$, the second equation of (\ref{fgpi1}) becomes
\begin{eqnarray}
\label{expfgpi1}
\frac{k^2 \delta f_{\pi}(k)}{g_{\pi qq}(k)} = m (k^2 \delta \Pi_{\pi}^{(2)}(0) + k^4 \delta \Pi_{\pi}^{(4)}(0)
                  + \cdots ).
\end{eqnarray}
By replacing in the Feynman diagrams in $kernel~b$ one vertex $i \gamma_5 {\vec \tau}$ 
by $i g_{\pi qq}(k) \gamma_5 \gamma_{\mu} {\vec \tau}/2$, the obtained subleading
axial-vector matrix element has the following form in external momentum expansion: 
\begin{eqnarray}
\label{expfey}
\frac{k_{\mu} \delta f_{\pi}(k)}{ g_{\pi qq}(k)} =k_{\mu} m(M(0) + k^2 M^{(2)}(0) + \cdots).
\end{eqnarray}
Comparing the two equations (\ref{expfgpi1}) and (\ref{expfey}), we derive the pion
polarization fluctuation in terms of the first coefficient of the expansion of 
the axial-vector matrix element.
\begin{eqnarray}
\label{compar}
\delta \Pi_{\pi}^{(2)}(0)=M(0)=M^{(b)}(0) + M^{(c)}(0) + M^{(d)}(0)
\end{eqnarray}
with
\begin{eqnarray}
\label{mbcd}
&M^{(b)}(0) & = i~N_c \{\int\frac{d^4q}
                {(2\pi)^4}[-{\rm i}D^{(RPA)}_{\pi}(q)](-3q^2L(q)) \nonumber \\
            &  & + \int\frac{d^4q}{(2\pi)^4}[-{\rm i}D^{(RPA)}_{\sigma}(q)]
              (4K(q)+3(4m^2-q^2)L(q)) \}, \\
&M^{(c)}(0) & = i~N_c \{ 6 \int\frac{d^4q}
   {(2\pi)^4}[-{\rm i}D^{(RPA)}_{\pi}(q)](K(q)+3K(0)-3q^2M(q)) \nonumber \\
            &  & +  2 \int\frac{d^4q}{(2\pi)^4}[-{\rm i}D^{(RPA)}_{\sigma}(q)] 
              (5K(q)+3K(0)-3(q^2-4m^2)M(q)) \}, \\
&M^{(d)}(0) & = -32iN_c~\int\frac{d^4q}{(2\pi)^4}
  N_c [-{\rm i}D^{(RPA)}_{\pi}(q)][-{\rm i}D^{(RPA)}_{\sigma}(q)] 
\{-(I(q)+2m^2K(0))(I(q)-I(0)) \nonumber \\
& & + (I(q)+I(0)-(q^2+2m^2)K(q))I(q) 
-q^2I(q)(I(q)+2m^2K(0))[-{\rm i}D^{(1)}_{\sigma}(q)] \},
\end{eqnarray}
where $L(q)$, $M(q)$ are defined in Appendix. In calculating $M^{(d)}(0)$, we 
have expanded the internal $\sigma$ propagator as
\begin{eqnarray}
\label{expsi}
[-{\rm i}D^{(RPA)}_{\sigma}(q-k)] = [-{\rm i}D^{(RPA)}_{\sigma}(q)](1+q_{\nu}k^{\nu}[-{\rm i}D^{(1)}_{\sigma}(q)] 
                           [-{\rm i}D^{(RPA)}_{\sigma}(q)]), 
\end{eqnarray}
with $
[-{\rm i}D^{(1)}_{\sigma}(q)] = 8N_c(I(q)+((4m^2-q^2)/2q^2)(I(q)-I(0)+q^2K(q))
          [-{\rm i}D^{(RPA)}_{\sigma}(q)])$.

\subsection{Internal RPA Meson Propagator}
~Till now, we have assumed that $kernel ~a$ and $kernel ~b$ shown in Fig. 2 are expanded
properly to leading and subleading order in $N_c$. 
In the chiral limit beyond mean-field approximation,
the leading meson 
polarization function $\Pi^{(RPA)}_{M}$ has the order of 
$O(N_c)$, which gives the RPA meson propagator $D^{(RPA)}_{M}(q)$ in the 
order of $O(1/N_c)$, thus $\delta m$ is the subleading $O(1/N_c)$ order of quark self-energy.
However, it is not the case when current quark mass is introduced
beyond mean-field approximation.
 
Now we first analyze the one-quark-loop pion propagator $D^{(RPA)}_{\pi}(q)$ 
beyond mean-field approximation and including $m_0$ explicitly.
The definition of RPA pion propagator is
\begin{eqnarray}
\label{rpapro1}
-{\rm i} D^{(RPA)}_{\pi}(q) = \frac{2iG}{1-2G \Pi^{(RPA)}_{\pi}(q)}.
\end{eqnarray}
Substituting  the expression of $\Pi^{(RPA)}_{\pi}(q)$, and using the gap equation 
Eq.~(\ref{gap}), the denominator is 
\begin{eqnarray}
\label{denom}
1-2G \Pi^{(RPA)}_{\pi}(q)=\frac{m_0+ \delta m}{m}+i~8GN_cq^2I(q).
\end{eqnarray}
With the pion pole condition Eq.~(\ref{pole}), we can deduce   
\begin{eqnarray}
\label{m0ovm}
\frac{m_0+ \delta m}{m}= -i~8GN_cm_{\pi}^2I(m_{\pi})+2G \delta \Pi_{\pi}(m_{\pi}).
\end{eqnarray} 
So the RPA pion propagator (\ref{rpapro1}) can be expressed as
\begin{eqnarray}
\label{rpapro2}
-{\rm i} D^{(RPA)}_{\pi}(q) = \frac{1}{4N_c(-m_{\pi}^2I(m_{\pi})
                       +q^2I(q))- i \delta \Pi_{\pi}(m_{\pi})}.
\end{eqnarray} 
Here $\delta \Pi_{\pi}(m_{\pi})$ is the subleading diagrams' contribution, which is 
in the order of $O(1)$. This term induces the RPA propagator expanded 
to $O(1/N_c^n)$ order, where $n$ could go to $\infty$. 
If we substitute this complete form of internal pion propagator into $kernel~a$, 
it is clear
that the $O(1/N_c)$ part of RPA propagator corresponds to the $O(1/N_c)$ part of $\delta m$, 
and other suppressed parts of RPA propagator correspond to the suppressed parts of $\delta m$.
 
In order to keep the proper subleading order $O(1/N_c)$ of $\delta m$, the internal 
meson propagator must be in its leading order $O(1/N_c)$.
We neglect the contributions resulted from $\delta \Pi_{\pi}(m_{\pi})$, and get 
the leading order of pion propagator in $O(1/N_c)$, 
\begin{eqnarray}
\label{rpapi}
-{\rm i} D^{(RPA)}_{\pi}(q) = \frac{1}{4N_c(-m_{\pi}^2I(m_{\pi})+q^2I(q))}.
\end{eqnarray}
Similarly, the $O(1/N_c)$ sigma propagator is
\begin{eqnarray}
\label{rpasig}
-{\rm i} D^{(RPA)}_{\sigma}(q) = \frac{1}{4N_c(-m_{\pi}^2I(m_{\pi})+(q^2-4m^2)I(q))}.
\end{eqnarray}
It can be seen that $G$ is not included explicitly in the two RPA meson propagators, and this 
simplifies our numerical calculations.

\section{Numerical results and discussion}
~Now we turn to the numerical evaluation. 
As evaluated in the last two chapters, we have three equations, 
the gap equation, Eq. (\ref{gap}), pion pole 
condition, Eq. (\ref{pole}), and pion
decay constant, Eq. (\ref{onshell}).
I the chiral limit, pion pole
condition Eq. (\ref{pole}) is trivial and the pion decay constant satisfies
the Goldberger-Treiman relation.

In order to deal with the divergence resulted from quark-loop integral and meson-loop
integral, like in \cite{Ann}, we introduce two cut-off, the quark momentum cut-off 
$\Lambda_f$ in Pauli-Villars regularization
and the meson momentum cut-off $\Lambda_b$ in covariant regularization.
As pointed out in Introduction, there are four parameters, the
current quark mass $m_0$, coupling constant $G$, quark momentum cut-off $\Lambda_f$ and meson
momentum cut-off $\Lambda_b$ to be fixed. By comparing with two observables $m_{\pi}=139 {\rm MeV}$,
$f_{\pi}=92.4 {\rm MeV}$ and one reasonable empirical range of
 $-300 {\rm MeV} < 1/2 <{\bar q}q >^{1/3} < -200 {\rm MeV}$,
we can not give fixed values of these four parameters. 
Here we regard the ratio $z= \Lambda_b / \Lambda_f$
as one free parameter. For each $z$, we can get a series of solutions
from the above conditions. The meson cloud effect is now characterized by $z$, the larger $z$
means more meson contributions. Specially,
when $z=0$, i.e., $\Lambda_b=0$, it returns to the
mean-field approximation.

In the case of explicit chiral symmetry breaking $m_0 \not= 0$, giving a $z$, for different $m$, 
we find a series of $\Lambda_f$ and
$m_0$ from the above  
three equations, and then $G$, $<{\bar q q}>$ and other quantities could be calculated.
In the case of chiral limit $m_0=0$, giving a $z$, for different $m$, we find 
$\Lambda_f$ and $G$ from Goldberger-Treiman
equation and the gap equation, then $<{\bar q q}>$ and other quantities
are calculated.  

In both cases, the mesonic contributions to the quark self-energy $m$ and to the pion
polarization $\Pi_{\pi}(k)$ are negative, and all the ratios of the mesonic corrections 
to the one-quark-loop
contributions to $m$, $\Pi_{\pi}(k)$ and physical quantities $f_{\pi}$ and quark condensate
increases with increasing $z$, and can reach 30-40$\%$ for values of parameters in the range 
$0.3{\rm GeV} < m < 0.5 {\rm GeV}$ and $ 1 < z < 1.5 $. These results are 
qualitatively the same as those
in \cite{Ann} and \cite{NPA}.

Our numerical results are shown in Figs. 3-5, which concentrate on
investigating the effect of explicit chiral symmetry breaking. In all figures, the 
solid lines
correspond to the case of explicit chiral symmetry breaking $m_0 \not= 0$, and 
the dashed lines
correspond to the case of chiral limit $m_0=0$, and $a, b, c, d, e$ 
correspond to $z=0, 0.5, 1, 1.5, 2,$
respectively. 

\begin{figure}[ht]
\centerline{\epsfxsize=10cm\epsffile{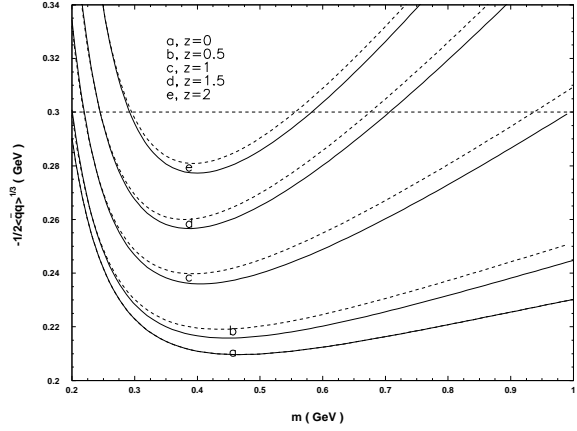}}
\caption
{Quark condensate $-1/2<{\bar q}q>^{1/3}$ as a function of the constituent quark mass $m$ for 
different values of 
$z=\Lambda_b/\Lambda_f$. The solid lines and dashed lines correspond
 to $m_0 \not=0$ and $m_0=0$
respectively.}
\label{mcond_fig}
\end{figure}

In Fig. \ref{mcond_fig}, we plot the quark scalar condensate $(-1/2<{\bar q}q>^{1/3})$
as a function of constituent quark mass $m$ for different values of $z$.  
This figure is different from the Fig. 6 and Fig. 7 in \cite{NPA}.  
Our numerical results show that: \\
~~~1). For each curve, the quark condensate has a plateau, in which 
the quark condensate changes slowly with $m$, and the width of the plateau 
becomes more and more narrow with increasing $z$. For each curve, we define 
the plateau in 
the range of $(-1/2<{\bar q}q>^{1/3})<(-1/2<{\bar q}q>^{1/3}_{min} + 0.0015)$ GeV.
\\
~~~2). For each $z$, the plateau of quark condensate is 
in the experimental range 0.2 GeV $\sim$ 0.3 GeV, and the plateau
becomes higher and higher with increasing $z$, which shows that the meson contributions affect 
the quark condensate explicitly. \\
~~~3). The solid and dashed lines of $a$ are almost coincide, which means that 
the effect of current quark mass is invisible in the mean-field approximation. 
While beyong mean-field approximation, i.e., considering the
meson contributions, the two curves in the case of $b$ or $c,d,e$ 
are separate in the region of the plateau. 
In Table 1, corresponding to the plateau region 
$(-1/2<{\bar q}q>^{1/3})<(-1/2<{\bar q}q>^{1/3}_{min} + 0.0015) {\rm GeV}$, 
\begin{center}
\begin{picture}(460,240)(0,0)
\SetWidth{1}
\EBox(0,0)(460,240)
\Line(0,60)(460,60)
\Line(0,120)(460,120)
\Line(0,180)(460,180)
\Line(40,0)(40,180)
\Line(40,30)(460,30)
\Line(40,90)(460,90)
\Line(40,150)(80,150)
\Line(240,150)(320,150)
\Line(80,0)(80,240)
\Text(20,30)[]{z=1.5}
\Text(60,15)[]{$m_0 \not= 0$}
\Text(60,45)[]{$m_0=0$}
\Text(20,90)[]{z=1}
\Text(60,75)[]{$m_0 \not= 0$}
\Text(60,105)[]{$m_0=0$}
\Text(20,150)[]{z=0}
\Text(60,135)[]{$m_0 \not= 0$}
\Text(60,165)[]{$m_0=0$}
\Line(160,0)(160,240)
\Line(240,0)(240,240)
\Line(320,0)(320,240)
\Line(400,0)(400,240)
\Text(120,220)[]{-${1 \over 2}<{\bar q}q>^{1/3}$}
\Text(120,150)[]{0.2096+0.0015}
\Text(120,105)[]{0.2397+0.0015}
\Text(120,75)[]{0.2360+0.0015}
\Text(120,45)[]{0.2599+0.0015}
\Text(120,15)[]{0.2566+0.0015}
\Text(120,200)[]{(GeV)}
\Text(200,210)[]{m(GeV)}
\Text(200,150)[]{0.47$\mp$0.1}
\Text(200,105)[]{0.39$\mp$0.05}
\Text(200,75)[]{0.40$\mp$0.05}
\Text(200,45)[]{0.38$\mp$0.04}
\Text(200,15)[]{0.39$\mp$0.04}
\Text(280,210)[]{$m_0$(MeV)}
\Text(280,135)[]{8.9$\mp$0.30}
\Text(280,75)[]{7.78$\mp$0.20}
\Text(280,15)[]{7.29$\mp$0.20}
\Text(360,210)[]{$\Lambda_f$(GeV)}
\Text(360,150)[]{0.615$\pm$0.002}
\Text(360,75)[]{0.71$\pm$0.01}
\Text(360,105)[]{0.725$\pm$0.01}
\Text(360,15)[]{0.785$\pm$0.008}
\Text(360,45)[]{0.801$\pm$0.008}
\Text(430,210)[]{G$\Lambda_f^2$}
\Text(430,150)[]{4.8$\mp$0.9}
\Text(430,75)[]{4.7$\mp$0.5}
\Text(430,105)[]{4.8$\mp$0.5}
\Text(430,15)[]{5.2$\mp$0.5}
\Text(430,45)[]{5.3$\mp$0.5}
\end{picture}
\end{center}
\begin{center}
{Table. 1. Quantities in the region of defined plateaus.}
\end{center} 
we list the
range of consituent quark mass $m$, current quark mass $m_0$, quark momentum cut-off $\Lambda_f$ and 
a dimensionless quantity $G \Lambda^2_f$.This table shows that, in the mean-field approximation, the values of constituent quark mass 
$m$ and the current mass $m_0$ in the plateau
is a little higher than the empirical values $m \simeq 1/3 ~proton ~mass$ and $m_0 \simeq 5\sim 7$MeV,
and the quark condensate in the plateau is much lower
than 0.25 GeV; however, these quantities in the plateaus at $z=1, 1.5$ are more reasonable 
comparing with the empirical values.

\begin{figure}[ht]
\centerline{\epsfxsize=10cm\epsffile{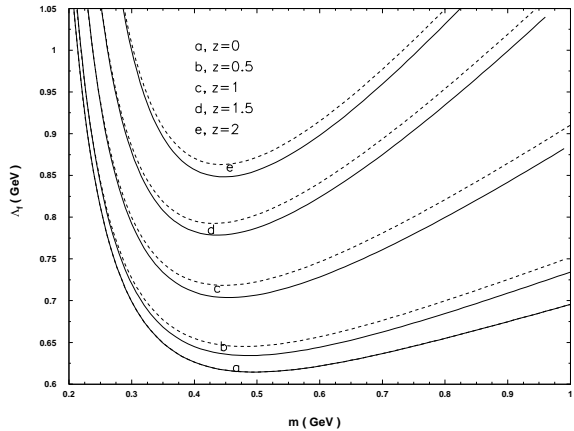}}
\caption
{Quark momentum cut-off $\Lambda_f$ as a function of the constituent quark mass $m$ for 
different values of 
$z=\Lambda_b/\Lambda_f$. The solid lines and dashed lines correspond
 to $m_0 \not=0$ and $m_0=0$
respectively.}
\label{mlf_fig}
\end{figure}

\begin{figure}[ht]
\centerline{\epsfxsize=10cm\epsffile{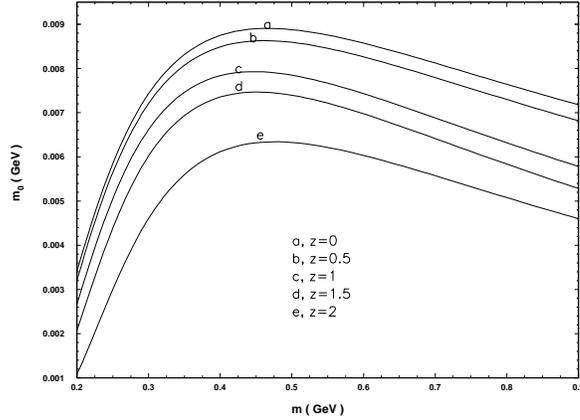}}
\caption
{Current quark mass $m_0$ as a function of the constituent quark mass $m$ for 
different values of 
$z=\Lambda_b/\Lambda_f$. The solid lines and dashed lines correspond
 to $m_0 \not=0$ and $m_0=0$
respectively.}
\label{mm0_fig}
\end{figure}

In Fig. \ref{mlf_fig}, we show the quark momentum cut-off $\Lambda_f$ 
as a function of the constituent quark mass $m$,  
\noindent and in Fig. \ref{mm0_fig} we show the current quark
mass $m_0$ as a function of the constituent quark mass $m$ for different values of $z$.
There is also a plateau for each curve. Although
the ranges of constituent quark mass $m$ corresponding to the $\Lambda_f$ plateau, $m_0$ 
plateau and quark condensate plateau are not exactly
the same, they are mostly around $m=0.4 \pm 0.1 {\rm GeV}$. This range $m=0.4 \pm 0.1{\rm GeV}$
is determined dominantly by the gap equation in the mean-field approximation, 
the effects of meson
cloud contributions are also shown clearly. Comparing the plateaus
at $z=0$ and $z=1.5$,
we see that $\Lambda_f$ and $(-1/2<{\bar q}q>^{1/3})$ are all modified by the
order of about $30 \%$. But the corresponding plateaus at $z=2$ modified 
too much. In fact,
all these figures show that the plateaus of $z=2$ are abnormal comparing that 
of $z=0.5, 1,1.5$. Based on the above analysis, it seems that  
$z=2$ is not a reasonable choice, and $z=1 \sim 1.5$ is more reasonable. 
In the region $m=0.4 \pm 0.1$ GeV, all these three figures 
Figs. \ref{mcond_fig}-\ref{mm0_fig} show that: \\
~1) In the mean-field approximation, the solid line and dashed line of $z=0$
coincide, which means that the effects of the current quark mass are invisible in this case. \\
~2) When beyond mean-field approximation $z>0$, i.e, considering meson 
contributions, the effects of current quark mass
are explicitly reflected by the separation of the solid and dashed lines. \\ 
As we pointed out in Introduction, in NJL model all information is calculated from constituent
quark propagator, and the current quark mass $m_0$ induces the massive pion $m_{\pi}$. 
In the mean-field approximation, 
there is no feedback of meson modes to quark propagator, so the effects of $m_0$ or $m_{\pi}$
can not be reflected. Only beyond mean-field approximation, when the 
back interaction of meson mode to quark propagator are considered, 
can the effects of $m_0$ or $m_{\pi}$ be reflected on the quantities
calculated from the fundamental element, quark propagator.

\section{Conclusions and Summary}
~We establish a general scheme of SU(2) NJL model including the
current quark mass explicitly and considering the meson cloud contributions.
To ensure the quark self-energy expanded in the proper order of $1/N_c$,
we discuss in details that the internal meson propagator must be in
$O(1/N_c)$ order. And we also develop a unified way to calculate all
Feynman diagrams in the NJL model by only expanding the constituent quark propagator 
in external momentum $k$. Our numerical results show that: a) when $z=1 \sim 1.5$, 
the NJL parameters are more reasonable than that in the 
mean-field approximation; b) the effect of current quark mass can  be 
reflected only beyond mean-field approximation, and it is characterized explicitly 
by the separation of the solid and dashed lines in the range of the parameters corresponding
to $m$ between $0.3 {\rm GeV} \sim 0.5 {\rm GeV}$.

\vskip 1cm
\begin{center}
{\bf  ACKNOWLEDGMENTS}
\end{center}
~~~The authors would like to thank  
Dr. V.~Dmitra{\v s}inovi{\' c}, 
Dr. E.N.~Nikolov and Dr. M. Franz for their kind help
during this work. This work is supported by NNSF of China
(Nos. 19677102 and 19845001). 

\newpage

\appendix
\begin{center}
{\bf APPENDIX: Definition of Integral Functions \\
and Corresponding Pauli-Villars Expressions}
\end{center}
\setcounter{equation}{0}
~The fundamental functions are defined as following:
\begin{eqnarray}
F=\int\frac{d^4p}{(2\pi)^4}\frac{1}{p^2-m^2},
\end{eqnarray}
\begin{eqnarray}
I(q)=\int\frac{d^4p}{(2\pi)^4}\frac{1}{(p^2-m^2)((p+q)^2-m^2)},
\end{eqnarray}
\begin{eqnarray}
K(q)=\int\frac{d^4p}{(2\pi)^4}\frac{1}{(p^2-m^2)^2((p+q)^2-m^2)},
\end{eqnarray}
\begin{eqnarray}
L(q)=\int\frac{d^4p}{(2\pi)^4}\frac{1}{(p^2-m^2)^2((p+q)^2-m^2)^2},
\end{eqnarray}
\begin{eqnarray}
M(q)=\int\frac{d^4p}{(2\pi)^4}\frac{1}{(p^2-m^2)^3((p+q)^2-m^2)},
\end{eqnarray}
And their corresponding Pauli-Villars expressions are:
\begin{eqnarray}
F_{PV}  =  -\frac{{\rm i}}{(4\pi)^2}\sum_{a}C_a{M_a}^2 \ln\frac{M_a^2}{m^2}, 
\end{eqnarray}
\begin{eqnarray}
I(q)_{PV}  =  \frac{{\rm i}}{(4\pi)^2}\sum_{a}C_a (-\ln\frac{M_a^2}{m^2} 
                +2(1-\sqrt{1+y_a^{-1}})\ln(\sqrt{y_a}+\sqrt{1+y_a}),
\end{eqnarray}
\begin{eqnarray}
K(q)_{PV}  =  -\frac{{\rm i}}{(4\pi)^2}\sum_{a}C_a \frac{1}{2M_a^2}\frac{
             \ln(\sqrt{y_a}+\sqrt{1+y_a})}{\sqrt{y_a(1+y_a)}},
\end{eqnarray}
\begin{eqnarray}
L(q)_{PV}  =  \frac{{\rm i}}{(4\pi)^2}\sum_{a}C_a (-\frac{1}{8M_a^4y_a(1+y_a)})
                (1-\frac{1+2y_a}{\sqrt{y_a(1+y_a)}}\ln(\sqrt{y_a}+\sqrt{1+y_a})),
\end{eqnarray}
\begin{eqnarray}
M(q)_{PV}  =  \frac{{\rm i}}{(4\pi)^2}\sum_{a}C_a (\frac{1}{16M_a^4y_a(1+y_a)})
                (1+2y_a-\frac{\ln(\sqrt{y_a}+\sqrt{1+y_a})}{\sqrt{y_a(1+y_a)}},
\end{eqnarray}
where $C_a=[1,1,-2]$, $\alpha_a=[0,2,1]$, $M_a^2=m^2(1+\alpha_a x)$ for a=0,1,2,
and $y_a=-q^2/(4M_a^2)$.

\end{document}